\documentclass[reprint,superscriptaddress,preprintnumbers,
groupedaddress,
nofootinbib,
 amsmath,amssymb,
 aps]{revtex4-1}
\usepackage{multirow}
\usepackage{upgreek}
\usepackage{tensor}
\usepackage{float}
\usepackage{latexsym}
\usepackage{graphicx}
\usepackage{amssymb}
\usepackage{amsmath}
\usepackage{amsfonts}
\usepackage[hidelinks]{hyperref}
\usepackage{color}
\usepackage{cool}
\usepackage{dcolumn}
\usepackage{slashed}
\usepackage{mathrsfs}

\begin{document}
\preprint{TUM-HEP-1314/21}

\title{Fractional dark energy}

\author{Ricardo G. Landim}\email{ricardo.landim@tum.de}

\affiliation{Technische Universit\"at M\"unchen, Physik-Department T70,\\ James-Franck-Stra\text{$\beta$}e 1, 85748 Garching, Germany}


\date{\today}

\begin{abstract}
In this paper we introduce the fractional dark energy model, in which the accelerated expansion of the Universe is driven by a nonrelativistic gas (composed by either fermions or bosons) with a noncanonical kinetic term.  The kinetic energy is inversely proportional to the cube of the absolute value of the momentum for a fluid with an equation of state parameter  equal to minus one, and whose corresponding energy density mimics the one of the cosmological constant. In the general case,  the dark energy equation of state parameter (times three) is precisely the exponent of the momentum in the kinetic term. We show that this inverse momentum operator appears in fractional quantum mechanics  and it is the inverse of the Riesz fractional derivative. The  observed vacuum energy can be obtained through the integral of the Fermi-Dirac (or Bose-Einstein) distribution and the lowest allowed energy of the particles.   Finally, a possible thermal production and fate of fractional dark energy is investigated.
\end{abstract}

\maketitle

\section{Introduction}

A satisfactory theoretical explanation for  the current accelerated expansion of the Universe is an open question that indicates the necessity of new physics. Discovered in 1998 using type-Ia supernovae \cite{perlmutter1999,reiss1998}, the acceleration of the Universe is described by a fluid with negative pressure (dark energy, hereafter DE), whose simplest candidate  is a cosmological constant $\Lambda$. However, the observed value of the vacuum energy ($10^{-47}$ GeV$^4$) is extremely smaller than any estimate of the zero-point energy of all modes of a field up to a cutoff scale \cite{Weinberg:1988cp}.   The lack of a good explanation for the origin of $\Lambda$ and its smallness  leads to the search of alternative candidates, such as scalar or vector fields \cite{peebles1988,ratra1988,Frieman1992,Frieman1995,Caldwell:1997ii,Padmanabhan:2002cp,Bagla:2002yn,ArmendarizPicon:2000dh,Brax1999,Copeland2000,Vagnozzi:2018jhn,Koivisto:2008xf,Bamba:2008ja,Emelyanov:2011ze,Emelyanov:2011wn,Emelyanov:2011kn,Kouwn:2015cdw,Landim:2015upa,Landim:2016dxh,Banerjee:2020xcn}, metastable DE \cite{Szydlowski:2017wlv,Stachowski:2016zpq,Stojkovic:2007dw,Greenwood:2008qp,Abdalla:2012ug,Shafieloo:2016bpk,Landim:2016isc, Landim:2017kyz,Landim:2017lyq}, holographic DE \cite{Hsu:2004ri,Li:2004rb,Pavon:2005yx,Wang:2005jx,Wang:2005pk,Wang:2005ph,Wang:2007ak,Landim:2015hqa,Li:2009bn,Li:2009zs,Li:2011sd,Saridakis:2017rdo,Mamon:2017crm,Mukherjee:2016lor,Feng:2016djj,Herrera:2016uci,Forte:2016ben}, interacting DE \cite{Wetterich:1994bg,Amendola:1999er,Guo:2004vg,Cai:2004dk,Guo:2004xx,Bi:2004ns,Gumjudpai:2005ry,Yin:2007vq,Costa:2013sva,Abdalla:2014cla,Costa:2014pba,Landim:2015poa,Landim:2015uda,Costa:2016tpb,Marcondes:2016reb,Landim:2016gpz,Wang:2016lxa,Farrar:2003uw,micheletti2009,Yang:2017yme,Marttens:2016cba,Yang:2017zjs,Costa:2018aoy,Yang:2018euj,Landim:2019lvl,Vagnozzi:2019kvw}, and usage of extra dimensions \cite{dvali2000}.

DE has the unusual property that its pressure is negative, and thus given the unsolved theoretical issues related to the origin of DE, one may wonder if it can be described by a matter with new properties.  In this paper, we extend the thermodynamic approach presented in \cite{Lima:2004wf} and introduce the fractional dark energy (FDE) model (the name ``fractional'' will be clear in Sec. \ref{sec:fqm}), in which DE is formed by a gas of nonrelativistic particles with a noncanonical kinetic term: an inverse momentum term. The DE equation of state parameter $w$ is simply the power of the inverse momentum term and the resulting energy density ends up mimicking  the one of the cosmological constant. The observed vacuum energy  can be obtained from the  integral of the corresponding Fermi-Dirac (or Bose-Einstein) distribution  with an appropriate lower limit of integration, which is related to the minimum allowed  energy of a FDE particle. 
The energy spectrum with an inverse momentum term is shown  to come from an inverse momentum operator in fractional quantum mechanics (FQM). FQM is a generalization of QM that appears when Lévy-like quantum paths are used in the Feynman's path integral approach, rather than the usual Brownian-like
quantum paths \cite{Laskin:1999tf}. The corresponding Schr\"odinger equation and several standard QM problems were investigated in the light of FQM in refs. \cite{laskin2000fractional,Laskin:2002zz,guo2006some,bayin2012consistency,bayin2012comment,dong2007some,de2010fractional}. Additionally,  FQM and fractional calculus have been applied to    quantum cosmology \cite{Moniz:2020emn,Rasouli:2021lgy} and Newtonian gravity \cite{Giusti:2020kcv,Giusti:2020rul}. 

Starting from a thermodynamic description of fluids with negative pressure \cite{Silva:2002fi,Lima:2004wf} (Sec. \ref{sec:past}), we show in Sec. \ref{sec:p3} that a nonrelativistic particle with an inverse momentum kinetic term, whose power is three times the DE equation of state parameter, can give rise to an energy density that mimics the one of the cosmological constant. The   corresponding  Fermi-Dirac (or Bose-Einstein) integral can result in a very small value, thus helping explain the observed vacuum energy.  The inverse momentum term may arise from an inverse operator in FQM, as explained in Sec. \ref{sec:fqm}. A possible thermal production and fate of FDE is investigated in Sec. \ref{sec:origin}. Sec. \ref{sec:conclu} is reserved for conclusions.

We will use natural units $\hbar=c=1$ throughout the text, unless explicitly stated.

\section{Thermodynamics of a  fluid with negative pressure}\label{sec:past}   
In this section we review the thermodynamic description of DE, according to \cite{Silva:2002fi,Lima:2004wf}. The second law of thermodynamics with a null chemical potential can be written as \cite{Weinberg:1971mx}
\begin{equation}
  n T  d\sigma = d\rho -\frac{\rho+P}{n}dn\,,
\end{equation}
where $n$ is the number density, $T$ is the temperature, $\sigma$ is the entropy per number of particles, $\rho$ is the energy density and $P$ is the pressure. The expression above  can be converted in the following equation after using the continuity equation, the conservation of the particle number and the fact that $d\sigma$ is an exact differential \cite{Silva:2002fi}
\begin{equation}
   \frac{\dot{T}}{T}= \frac{\partial P}{\partial \rho}\frac{\dot{n}}{n}\,.
\end{equation}
Using a constant equation of state parameter $w\neq 0$  one can obtain \cite{Lima:2004wf}
\begin{equation}
    n\propto T^{\frac{1}{w}}\,,
\end{equation}
and since $n\propto V^{-1}$, where $V$ is the volume, we have
\begin{equation}\label{eq:TV_relation}
    T^{\frac{1}{w}}V=\text{const.}
\end{equation}
For $w=-1$ the temperature scales linearly with the volume. Using the continuity equation the energy density for DE is written as
\begin{equation}\label{eq:rho1}
    \rho\propto V^{-(1+w)}\,,
\end{equation}
where $V\propto a^3$ and $a$ is the scale factor. If  Eq. (\ref{eq:TV_relation}) is used in Eq. (\ref{eq:rho1}) the following relation is obtained
\begin{equation}\label{rho_T0}
    \rho\propto T^{\frac{1+w}{w}}\,.
\end{equation}
   The equation above, in turn, can be obtained from
   \begin{equation}\label{eq:rhoint}
       \rho=C_0\int_0^\infty\frac{\varepsilon^{\frac{1}{w}}}{e^{\beta \varepsilon}+1}d \varepsilon\,,
   \end{equation}
    where $\beta=(k_B T)^{-1}$, $k_B$ is the Boltzmann constant (set to one hereafter) and $C_0$ is a constant. Although it was not addressed in \cite{Lima:2004wf}, here we consider only fermions because  for bosons the integral can be performed through analytic continuation,  however, the result is negative \cite{cvijovic2009fermi}, yielding therefore a negative energy density.
    
    Finally, the entropy for a DE candidate in this approach is \cite{Lima:2004wf}
    \begin{equation}
        S\propto (1+w)T^{1/w}V\,,
    \end{equation}
    implying that a phantom DE candidate ($w<-1$) is not thermodynamically allowed.\footnote{Although other  models may allow  phantom cosmology from a thermodynamic point of view \cite{Cruz:2018arw}.}
    
    \section{Dark energy as a nonrelativistic  gas with a noncanonical kinetic term}\label{sec:p3}

As mentioned in the last section,  bosons  provide a negative energy density in the equivalent Eq. (\ref{eq:rhoint}) and this potential issue was not covered in \cite{Lima:2004wf}. Furthermore, neither relativistic nor nonrelativistic particles obeying a canonical energy-momentum relation would yield a density of states $D(\varepsilon
)\propto \varepsilon^{1/w-1}$, thus one may wonder whether  Eq. (\ref{eq:rhoint}) can be obtained  from arguments other than the ones presented in \cite{Lima:2004wf}. In fact, since the density of states $D(\varepsilon
)\propto \varepsilon^{1/w-1}$ is noncanonical and it is divergent for $w<0$ and $\varepsilon\rightarrow 0$, it is a fair assumption that the energy-momentum relation that describes DE is noncanonical as well, with a possible divergence. 

In order to qualitatively investigate  the above-mentioned arguments,  we start the discussion considering a gas of fermions, but as it will be clear below, the same procedure and results apply to bosons as well.

The integral in Eq. (\ref{eq:rhoint}) can be evaluated in the following way.
   The Fermi-Dirac integral \cite{dingle1957fermi}
    \begin{equation}\label{eq:FDint}
        \mathscr{F}_a(z)= \frac{1}{\Gamma (a+1)}\int_0^\infty \frac{t^a}{e^{t-z}}dz\quad a>-1\,, z \in \mathbb{R}\,,
    \end{equation}
   where $\Gamma(a+1)$ is the Gamma function, can be evaluated for negative powers $a\leq -1$ using the  relation \cite{goano1995algorithm}
   \begin{equation}\label{eq:FD_integral_deriv}
    \frac{d}{dz}\mathscr{F}_a(z)   = \mathscr{F}_{a-1}(z)\,.
   \end{equation}
   Therefore, for $a=-1$ we have \cite{goano1995algorithm,cvijovic2009fermi}
   \begin{equation}
       \mathscr{F}_{-1}(z) =\frac{1}{1+e^{-z}}\,,
   \end{equation}
   which gives $1/2$ if $z=0$, for example. 
   
   As the energy density is $\rho(\varepsilon)=\varepsilon D(\varepsilon)f(\varepsilon)$, where $f(\varepsilon)$ is the Fermi-Dirac distribution, the density of states $D(\varepsilon)\propto\varepsilon^{\frac{1}{w}-1}$ is obtained if the energy spectrum of a  Fermi gas has an additional noncanonical kinetic term
    \begin{equation}\label{eq:EMrelation}
       \varepsilon^2=p^2+m^2+\frac{C^2}{p^{-6 w}}\,,
   \end{equation}
   where $p\equiv |\mathbf{p}|$ is the three-momentum and $C$ is a ``coupling'' constant with dimensions of  [energy]$^{1-3w}$. Eq. (\ref{eq:EMrelation})  can be approximately expanded  for a nonrelativistic gas ($p^2\ll m^2$) as
   \begin{equation}\label{eq:EMrelation2}
       \varepsilon\approx m+\frac{p^2}{2 m}+\frac{C}{p^{-3 w}}\,,
   \end{equation}
   where  a canonical nonrelativistic matter is recovered when $C\rightarrow 0$. This exotic matter with a modified energy-momentum relation can  describe DE only if  $Cp^{3w}\gg p^2/(2m)$ (or rather $Cp^{3w}\gg m$), yielding
       \begin{equation}\label{eq:energy_p3}
       \varepsilon\approx \frac{C}{p^{-3 w}}\,.
   \end{equation}
The momentum scales with $a^{-1}$, and thus the energy for a fluid with $w=-1$ scales with $a^3$, as shown in \cite{Lima:2004wf}.

Different from canonical nonrelativistic or relativistic particles, not all values of energy are allowed for the FDE particle. Before proceeding to the explicit calculation of the DE energy density and number density, we have to investigate what are those values of energy that enable DE to be described by Eq. (\ref{eq:energy_p3}).  Whatever  the mechanism that generates  FDE in the early Universe, it is expected that  FDE ``freezes out" when the particles are already nonrelativistic (according to Eq. (\ref{eq:EMrelation2})). More than that, the momentum not only should be  much smaller than the rest mass but also it should be $p\ll (C/m)^{-\frac{1}{3w}}$, in order for the approximation in Eq. (\ref{eq:energy_p3}) to be valid. Therefore, the maximum momentum $p_{\max}\lesssim(C/m)^{-\frac{1}{3w}}$ is equivalent to a minimum energy $\varepsilon_{\min}$ for FDE. 

On the other hand, in order to avoid a divergence in the energy when momentum approaches zero, one might expect that there is a minimum momentum, and therefore a maximum energy. This maximum energy will be described  later.

Now we can return to the calculation of the density of states, which is done similarly to the case of an ideal Fermi gas. The total number of particles is found integrating the particle distribution for the absolute magnitude of the momentum, written in spherical coordinates \cite{landau1980statistical}
\begin{equation}
    N_p=\frac{gV}{2\pi^2}\int_{p_{\min}}^{p_{\max}} \frac{p^2 }{e^{\beta \varepsilon}+1}dp\,,
\end{equation}
where $V$ is the volume of the gas and $g=2s+1$ is the spin multiplicity. The limits of integration are therefore not zero and infinity as it was in Eq. (\ref{eq:rhoint}). Using Eq. (\ref{eq:energy_p3}) we can write the integral above in terms of the energy $\varepsilon$
\begin{equation}
    N_\varepsilon=-\frac{C^{-\frac{1}{w}}gV}{6\pi^2 w}\int_{\varepsilon_{\min}}^{\varepsilon_{\max}} \frac{\varepsilon^{\frac{1}{w}-1} }{e^{\beta \varepsilon}+1}d\varepsilon\,.
\end{equation}
Therefore, the number density is
\begin{align}
    n&=-\frac{C^{-\frac{1}{w}}g}{6\pi^2w}\int_{\varepsilon_{\min}}^{\varepsilon_{\max}} \frac{\varepsilon^{\frac{1}{w}-1} }{e^{\beta \varepsilon}+1}d\varepsilon\,\\
    &= -\frac{C^{-\frac{1}{w}}g}{6\pi^2w}\beta^{-\frac{1}{w}} \mathscr{F}^{u_{\max}}_{u_{\min}, \frac{1}{w}-1}\label{eq:n_de}\,,
\end{align}
and $\mathscr{F}^{u_{\max}}_{u_{\min}, \frac{1}{w}-1}$ is the result of the Fermi-Dirac integral with the  appropriate limits of integration
\begin{equation}
  \mathscr{F}^{u_{\max}}_{u_{\min,a}}\equiv \int_{u_{\min}}^{u_{\max}}\frac{u^{\frac{1}{a}} }{e^u+1}du\,,
\end{equation}
where $u\equiv\beta \varepsilon$. 
 The negative sign in the number density is not a problem because for DE $w< -1/3$. 

The energy density is then given by
\begin{align}
    \rho&=-\frac{C^{-\frac{1}{w}}g}{6\pi^2w}\int_{\varepsilon_{\min}}^{\varepsilon_{\max}} \frac{\varepsilon^{\frac{1}{w}} }{e^{\beta \varepsilon}+1}d\varepsilon\,\\
    &= -\frac{C^{-\frac{1}{w}}g}{6\pi^2w}\beta^{-\frac{1+w}{w}} \mathscr{F}^{u_{\max}}_{u_{\min}, \frac{1}{w}}\,,\label{eq:rho_final}
\end{align}
in agreement with Eq. (\ref{rho_T0}). The FDE energy density is constant for $w=-1$ and it can be written in terms of the number density as
\begin{equation}\label{eq:integra_u}
    \rho=\beta^{-1}\frac{\mathscr{F}^{u_{\max}}_{u_{\min},\frac{1}{w}}}{\mathscr{F}^{u_{\max}}_{u_{\min}},\frac{1}{w}}n\,.
\end{equation}

The Landau potential $\Omega$ for FDE is
\begin{align}
    \frac{\Omega}{V}&=\frac{C^{-\frac{1}{w}}g\beta^{-1}}{6\pi^2w}\int_{\varepsilon_{\min}}^{\varepsilon_{\max}}\varepsilon^{\frac{1}{w}-1} \ln (e^{-\beta \varepsilon}+1)d\varepsilon\,\nonumber\\
      & =\frac{w C^{-\frac{1}{w}} g\beta^{-1}}{6\pi^2w}\Bigg[\ln(e^{-\beta \varepsilon}+1)\varepsilon^\frac{1}{w}\Bigg |_{\varepsilon_{\min}}^{\varepsilon_{\max}}\nonumber\\&+\int_{ \varepsilon_{\min}}^{\varepsilon_{\max}} \frac{\beta\varepsilon^{\frac{1}{w}}}{e^{\beta \varepsilon}+1}d\varepsilon\Bigg ]\,\nonumber\\
      & =\frac{w C^{-\frac{1}{w}} g}{6\pi^2w}\int_{ \varepsilon_{\min}}^{\varepsilon_{\max}} \frac{\beta\varepsilon^{\frac{1}{w}}}{e^{\beta \varepsilon}+1}d\varepsilon\,\nonumber\\
   &=-w\rho\,,
\end{align}
where in the second line we integrated by parts and the first term vanishes because $\varepsilon_{\max}\gg \varepsilon_{\min}$, thus $ \varepsilon_{\max}^\frac{1}{w}\approx 0$, and $\beta \varepsilon_{\min}\gtrsim m/T\gg 1$ thus $\ln(e^{-\beta \varepsilon_{\min}}+1)\approx 0$. The lower limit of integration cannot be zero because it would cause a divergence in the integral, therefore showing that the original treatment presented in \cite{Lima:2004wf} would suffer from this issue.
Since $\Omega=-PV$, where $P$ is the fluid pressure, we get the DE equation of state $P=w\rho$, as it should be.

From Eq. (\ref{eq:rho_final}) it is possible to obtain the DE energy density for spin-1/2 particles and $w=-1$, for instance. The value $\mathscr{F}^{u_{\max}}_{ u_{\min},-1}$ can be found numerically and it depends on both minimum and maximum energy. Given that  $u_{\min}$ in Eq. (\ref{eq:integra_u}) is $m/T\gg 1$, we can expect that the minimum value of $u$ is $10$ or $100$ for example. On the other hand, the maximum energy can in principle be arbitrary. It is possible then to obtain the result of the integral considering different limits of integration. For $u_{\min}=10$ and any arbitrary $u_{\max}\gtrsim 10 \,u_{\min}$ we obtain $\mathscr{F}^{\gtrsim 10^2 }_{10, -1}\approx 4\times 10^{-6}$, while if $u_{\min}=10^2$, $\mathscr{F}^{\gtrsim 10^3}_{ 10^2, -1}\approx 3\times 10^{-46}$. A larger $u_{\min}$ gives a much smaller result $\mathscr{F}^{\gtrsim 10^4}_{ 10^3, -1}\approx 5\times 10^{-438}$. These results show that in order to obtain the DE energy density $10^{-47}$ GeV$^4$, the constant $C$ can be of order $\mathcal{O}(1)  \text{ GeV}^4$, if the minimum energy per temperature is 100. The observed value of the cosmological constant  is then translated to the  minimum  allowed value of $\beta \varepsilon$.

Although we have done the procedure described here for fermions, the idea can be applied to bosons as well because the minimum (and maximum) energy per temperature is larger than one, thus $e^{\beta \varepsilon}\gg 1$ and the distribution describes a classical gas. Since the quantum statistics is suppressed, the results for bosons and fermions are  virtually the same.

\begin{figure*}
    \centering
    \includegraphics[scale=0.5]{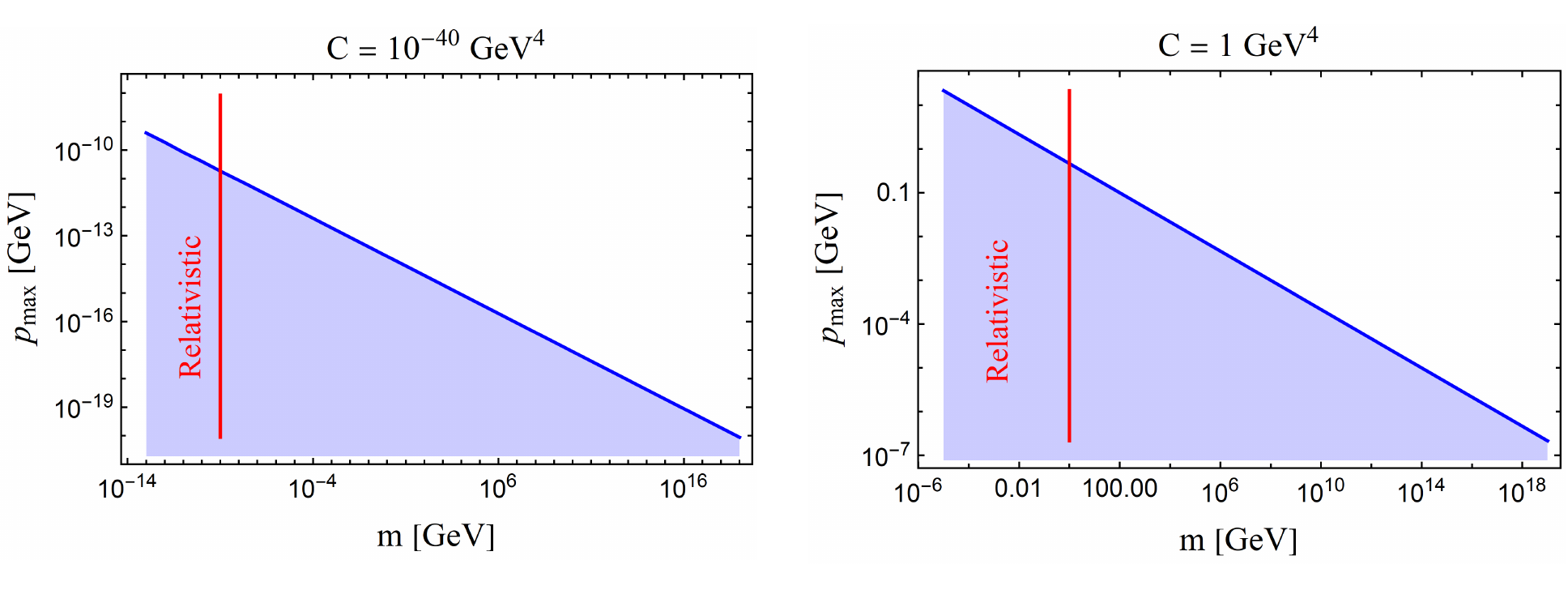}
    \caption{Maximum momentum (blue line) as a function of the particle mass $m$, for $u_{\min}=10$ (left)  and $u_{\min}=100$ (right). The light blue region shows when Eq. (\ref{eq:energy_p3}) is valid and the region to the right of the red line presents the regime  when the particles are nonrelativistic.}
    \label{fig:p_max}
\end{figure*}

In order to investigate the maximum momentum that allows Eq. (\ref{eq:energy_p3}) to be true, we use  $\frac{C/p^3}{m}=10$, thus $p_{\max}=[ C/(10m)]^{1/3}$, and we investigate two cases: $C=10^{-40}$ GeV$^4$ (for $u_{\min}=10$) and $C=1$ GeV$^4$ (for $u_{\min}=100$). The results are presented in Fig. \ref{fig:p_max}, where we also show the limit where the particles are nonrelativistic ($p_{\max}\ll m$). A slightly larger ratio $\frac{C/p^3}{m}=30$ gives very similar results.

\section{Fractional Quantum Mechanics}\label{sec:fqm}

One may wonder if there is a quantum mechanical operator that could give the energy relation (\ref{eq:energy_p3}) with the noncanonical kinetic term. The term $|\mathbf{p}|^{3w}$ should arise from an inverse Laplace operator $\Delta^{3w/2}$ (since $w$ is negative), where $\Delta\equiv \nabla^2$ is the three-dimensional (3D) Laplacian. A fractional Laplacian is defined in fractional calculus (see, for example \cite{podlubny1998fractional,oldham2006fractional}) and in the so-called FQM \cite{Laskin:1999tf,laskin2000fractional,Laskin:2002zz,laskin2018fractional}.
In this section we will use explicitly  the reduced Planck constant $\hbar$ and the absolute value of the 3D momentum $|\mathbf{p}|$. 

A fractional derivative can be defined as the Riemann-Liouville derivative $_aD^\alpha_b$ of order $\alpha$ \cite{oldham2006fractional}
\begin{equation}
    _aD^\alpha_bf(x)=\frac{1}{\Gamma(n+1-\alpha)}\frac{d^{n+1}}{dx^{n+1}}\int^x_a(x-y)^{n-\alpha}f(y)dy\,, 
\end{equation}
for $n\leq \alpha<n+1$, where $\Gamma(n+1-\alpha)$ is the Gamma function.
The Riemann-Liouville derivative 
operator is a left inverse of the Riemann-Liouville fractional integration
operator of the same order $\alpha$, that is, $_aD^\alpha_b(_aD^{-\alpha}_bf(x))=f(x)$. The inverse of the fractional derivative can be defined as \cite{oldham2006fractional}
\begin{equation}
    _aD^{-\alpha}_bf(x)=\frac{1}{\Gamma(\alpha)}\int^x_a(x-y)^{\alpha-1}f(y)dy\,, \quad  \alpha>0.
\end{equation}

The fractional derivatives and integrals have the important property $_aD^{\pm\alpha}_b(_aD^{\pm\beta}_bf(x))= {}_aD^{\alpha\pm\beta}_bf(x)$.

FQM was initially developed using Lévy trajectories in the path integral \cite{Laskin:1999tf} and resulted in the fractional Schrödinger equation in 3D (see \cite{laskin2018fractional} for a recent review) \cite{Laskin:1999tf,laskin2000fractals,Laskin:2002zz}
\begin{equation}
    i\hbar\frac{\partial \psi(\mathbf{r},t) }{\partial t}=A_\alpha  (-\hbar^2\Delta)^{\alpha/2}\psi(\mathbf{r},t)+V(\mathbf{r})\psi(\mathbf{r},t)\,,
\end{equation}
where $A_\alpha$ is a scale coefficient with units of $[A_\alpha]=\text{erg}^{1-\alpha}$cm$^\alpha$s$^{-\alpha}$, the momentum operator has the usual form $\mathbf{p}=-i\hbar \nabla$ and the Riesz fractional derivative \cite{riesz1949integrale} is given by \cite{Laskin:1999tf,laskin2000fractional}
\begin{equation}
    (-\hbar^2\Delta)^{\alpha/2}\psi(\mathbf{r},t)= \frac{1}{(2\pi \hbar)^3}\int d^3p e^{i \frac{\mathbf{p}\cdot \mathbf{r}}{\hbar}}|\mathbf{p}|^{\alpha} \varphi(\mathbf{p},t)\,,
    \end{equation}
with the  L\'evy index $\alpha$ lying  between $1< \alpha \leq 2$. 

Similarly to how  the Riesz fractional derivative in FQM  is defined \cite{laskin2018fractional}, we may define the inverse Riesz fractional derivative (or Riesz fractional integral) in 1-D, as
\begin{equation}
    \left(-i\hbar\frac{\partial}{\partial x}\right)^{-\alpha}\equiv \frac{1}{2}\left(  _{-\infty} D^{-\alpha}_x+  _xD^{-\alpha}_\infty\right)\,,
\end{equation}
which gives the result in 3D
\begin{equation}\label{eq:inverse_op}
    (-\hbar^2\Delta)^{-\alpha/2}\psi(\mathbf{r},t)= \frac{1}{(2\pi \hbar)^3}\int d^3p e^{i \mathbf{p}\cdot \mathbf{r}/\hbar}|\mathbf{p}|^{-\alpha} \varphi(\mathbf{p},t)\,.
    \end{equation}
It is clear that  the Riesz fractional integral is the inverse operator of the Riesz fractional derivative, as it should be, and  $e^{i\mathbf{p\cdot r}/\hbar}$ is the eigenfunction of the Riesz fractional integral operator, with eigenvalue $|\mathbf{p}|^{-\alpha}$.

Therefore, Eq. (\ref{eq:energy_p3}) determines the Hamiltonian operator for FDE, where in our case $\alpha=3w$ and the corresponding Schr\"odinger equation is given by
\begin{equation}
    i\hbar\frac{\partial \psi(\mathbf{r},t) }{\partial t}=C (-\hbar^2\Delta)^{3w/2}\psi(\mathbf{r},t)\,,
\end{equation}
where $C$ has units of $[C]= \text{erg}^{1-3w}$cm$^{3w}$s$^{-3w}$ (as presented before in natural units) and the inverse operator $(-\hbar^2\Delta)^{3w/2}$ is obtained applying twice the operator defined in Eq. (\ref{eq:inverse_op}), because the FQM is defined for  $1<\alpha \leq 2$. The result, $(-\hbar^2\Delta)^{3w/2}=(-\hbar^2\Delta)^{3w/4}(-\hbar^2\Delta)^{3w/4}$, is trivial because the eigenvalue of the operator $\Delta^{-(\alpha+\beta)/2} $ is $|\mathbf{p}|^{-(\alpha+\beta)}$. Similar to the Riesz fractional derivative, the Riesz fractional integral operator is Hermitian \cite{Laskin:2002zz} and the probability continuity
equation does not have a source term \cite{wei2016comment}, avoiding particle teleportation, since here we have a free particle with defined kinetic energy (excluding the region $|\mathbf{p}|\rightarrow 0$, as discussed before and as will be next).

\section{Origin and fate of FDE}\label{sec:origin}
While the understanding  of the origin  and fate of FDE from first principles is still open, we can investigate these phases of the particle's life in a first approximation, assuming $w=-1$ for simplicity. 

A canonical nonrelativistic particle has the classical energy $\varepsilon=m+p^2/(2m)$ when $m\gg C/p^3$, while DE is dominant when $m\ll C/p^3$; thus we may expect that the abundance in both regimes (nonrelativistic versus DE) are similar when $m\simeq C/p^3$. We can consider that in the usual nonrelativistic regime, FDE is described by a nonrelativistic fluid that froze out similarly to  dark matter.  It is then expected that the abundance of FDE $Y\equiv n/s$, where $s$ in this section is the entropy density, is roughly the same when $\varepsilon\sim m\sim C/p^3$. 

As it was seen in Sec. \ref{sec:p3}, the allowed minimum energy that gives the correct DE energy density today with $C$ of order of unity is $\varepsilon_{\min}=100 T=(10-30) m $. Since canonical nonrelativistic particles at the freeze-out have $x\equiv m/T\gtrsim 3$, the minimum energy per temperature $\varepsilon_{\min}/T$ is roughly $10-30$ times larger than $x$ at the freeze-out. Therefore, the nonrelativistic regime for FDE is around the corresponding value of $m/T\sim 3-10$ at the freeze-out.

In order to have a rough estimate of $Y$, we should calculate it in the two regimes and see if they match. The abundance of a canonical nonrelativistic particle at equilibrium, dark matter for instance, is \cite{kolb1981early}
\begin{equation}\label{eq:Yeq}
    Y_{\rm EQ}(x)=\frac{45}{2\pi^4}\left(\frac{\pi}{8}\right)^{1/2}\frac{g}{g\star}x^{3/2}e^{-x}\,,
\end{equation}
where $g\star$ is the effective number of relativistic species. Using Eq. (\ref{eq:n_de}), the abundance of a particle with the noncanonical kinetic term $C/p^3$ is
\begin{equation}\label{eq:Yde}
    Y(x)=\frac{45}{12\pi^4}\frac{g}{g*}\left(\frac{C}{m^4}\right)x^{4}\mathscr{F}^{100}_{3, -2}\,,
\end{equation}
where the limits in the integral $\mathscr{F}^{100}_{3, -2}$ were chosen to describe the nonrelativistic regime $3 \lesssim\varepsilon/T=m/T\lesssim 100$. 

\begin{figure}
    \centering
    \includegraphics[scale=0.5]{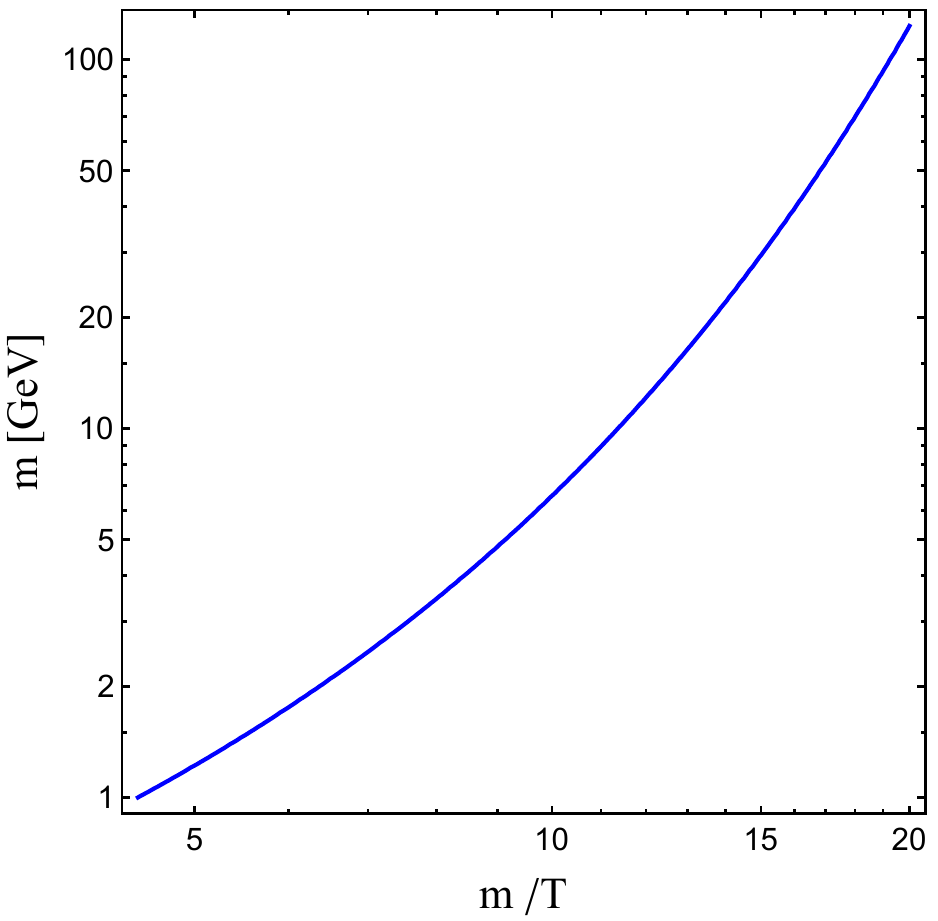}
    \caption{Parameter space in which the abundance of FDE in the nonrelativistic regime  ($\varepsilon\sim m$) is equal to the one in the  inverse momentum phase ($\varepsilon\sim C/p^3$). The particle is nonrelativistic for $m\gtrsim 1$ GeV.  }
    \label{fig:freezeout}
\end{figure}

Equating Eqs. (\ref{eq:Yeq}) and (\ref{eq:Yde}), with $C=1$ GeV$^4$ and $\mathscr{F}^{100}_{3, -2}=3\times 10^{-3}$, we obtain an estimate of the FDE mass and $m/T$ at the ``freeze-out'', shown in Fig. \ref{fig:freezeout}. When FDE becomes out-of-equilibrium  ($3\lesssim x_f\lesssim 10$) it does not freeze out in the usual sense (contrary to what happens to dark matter),  but its behavior changes from  nonrelativistic to the one dominated by the noncanonical kinetic term where it ``freezes'' out at $\varepsilon/T=100$. The range of masses that gives $3\lesssim x_f\lesssim 10$ is around $1 \text{ GeV}\lesssim m \lesssim 6 $ GeV. 

\begin{figure}
    \centering
    \includegraphics[scale=0.5]{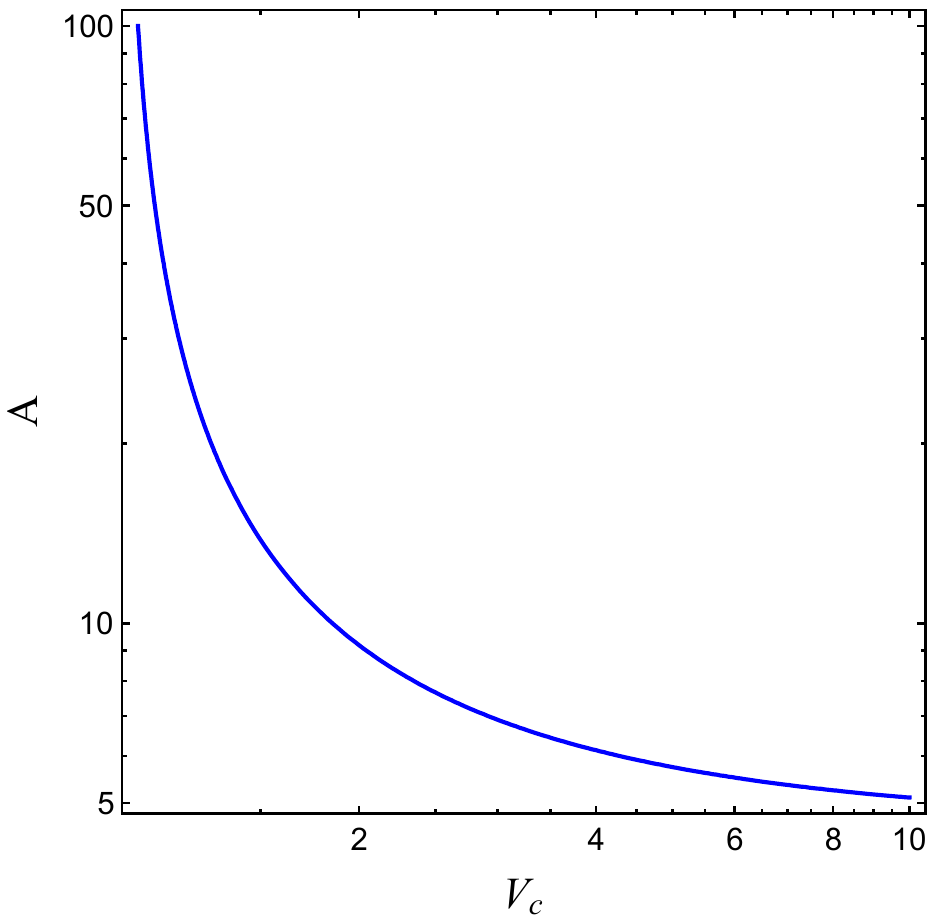}
    \caption{Parameters $A$ and $V_c$ from the phenomenological decay rate (\ref{eq:decay}), which satisfies the requirement of being within the errors of the DE equation of state parameter, such that  FDE has not decayed yet ($V=1$),   $Q/(3H_0)\sim 10^{-2}$.}
    \label{fig:Vc}
\end{figure}

Since  the temperature of FDE increases with the volume and $p\rightarrow 0$ is in principle problematic, one may expect that its behavior will not last forever, but will be converted somehow back to the nonrelativistic case, for example. We can consider that FDE is  long-lived but will decay into another nonrelativistic  particle $\phi$ in the future. The conservation of the energy-momentum tensor implies that the continuity equations are
\begin{align}
    \dot{\rho}_{\rm de}&=-Q \rho_{\rm de}\,,\\
    \dot{\rho}_{\phi}+3H\rho_{\phi}&=Q \rho_{\rm de}\,,
\end{align}
where $H$ is the Hubble rate and $Q$ is the decay rate. In order to parametrize this decay rate, we can use a phenomenological sigmoid function
\begin{equation}\label{eq:decay}
    Q=\frac{q}{1+e^{-A\frac{(V-V_c)}{V_c}}}\,,
\end{equation}
where $q$ is a constant with dimension of [time]$^{-1}$, $A>1$ is a dimensionless constant, $V$ is the  volume of the Universe and $V_c$ is a critical volume. When $V\ll V_c$ the decay rate is zero and when $V\gg V_c$ the sigmoid function is 1. We can  get an estimate on $A$ and $V_c$ by requiring that the decay rate is within the errors in the measurement of the DE equation of state parameter. If $q=3H_0$ and the volume of the Universe  is scaled to be 1 today, we can obtain the possible values of $A$ and $V_c$ that give $Q/(3H_0)\sim 10^{-2}$, which  is the error in the measurement of $w$ from \textit{Planck} \cite{Aghanim:2018eyx}. The result is shown in Fig. \ref{fig:Vc}.

The phenomenological decay (\ref{eq:decay}) assumes a dependence on the volume of the Universe, which means that after the critical volume the energy $\varepsilon=C/p^3$ reaches a maximum and stops increasing, thus decaying to another nonrelativistic particle. Since $V\sim T$ for $w=-1$, Eq. (\ref{eq:decay}) can be written in terms of the DE temperature, rather than the volume of the Universe. Then, when the FDE temperature reaches a critical value $T_c$, the decay is turned on. A more detailed explanation of this phenomenon is expected to come from first principles, and it is a subject of study in  future work.

\section{conclusions} \label{sec:conclu}
In this work we  have extended the thermodynamic approach of \cite{Lima:2004wf} and introduced the FDE model. The accelerated expansion of the Universe is  caused by a nonrelativistic gas with energy inversely proportional to the momentum to some power. The DE equation of state parameter  is precisely the exponent of the noncanonical kinetic term. When $w=-1$ the corresponding energy density of the gas mimics the one of the cosmological constant. The integration of the quantum distribution functions can result in a small enough value, such that the ``coupling'' constant $C$ of the modified Einstein energy-momentum relation can be of order $\mathcal{O}(1)$  GeV$^4$,  giving rise to the  observed value of the vacuum energy.

This noncanonical kinetic term is the eigenvalue of the inverse momentum operator, in the FQM framework. In this case, the operator is the inverse of the Riesz derivative, which in turn appears in the generalized Schr\"odinger equation. 

Because of the possible divergence when the momentum  goes to zero and since the inverse momentum should be much larger than the particle rest mass, there are corresponding maximum and minimum allowed energies for the FDE gas, respectively. If FDE particles are thermally generated in a similar fashion of the dark matter freeze out, then the parameter space can be constrained. Of course, if FDE is generated through a different phenomenon in the early Universe then the estimates made in Sec. \ref{sec:origin} do not hold. 

The same happens to the particle's maximum allowed energy. We investigated a phenomenological decay of the FDE particles into a new constituent of the dark sector, and this decay will start after a critical volume (or critical DE temperature). However,  DE's fate can be different if other mechanisms are evoked to avoid an infinite energy. Finally, it   remains open as a quantum field theory formulation of  FDE, which, along with the aforementioned points and potential signatures of FDE, will be further explored in a future work.

\acknowledgments
We thank the Alexander von Humboldt Foundation and  CAPES (process 88881.162206/2017-01) for the financial support.

\bibliography{references}

\end{document}